# Second-order stationary solutions for fermions in an external Coulomb field

V.P.Neznamov[1,2*], I.I.Safronov[1]

[1]RFNC-VNIIEF, Russia, Sarov, Mira pr., 37, 607188
[2]National Research Nuclear University MEPhI, Moscow, Russia

Abstract

We have studied self-conjugate second-order equations with spinor wavefunctions for fermions moving in an external Coulomb field. For stationary states, the equations are characterized by separated states with positive and negative energies, which render probabilistic interpretation possible. For the Coulomb field of attraction, the energy spectrum of the second-order equation coincides with the spectrum of the Dirac equation, while the probability densities of states are slightly different. For a Coulomb field of repulsion, there exists an impermeable potential barrier with radius depending on the classical electron radius and on the electron energy. The existence of the impermeable barrier does not contradict the results of experiment for determining the inner electron structure and does not affect (in the lowest order of perturbation theory) the Coulomb electron scattering cross section. The existence of the impermeable barrier can lead to positron confinement in supercritical nuclei with $Z \geq 170$ in case of realization of spontaneous emission of vacuum electron-positron pairs.

---

[*] E-mail: neznamov@vniief.ru, vpneznamov@mail.ru

## 1. Introduction

In quantum mechanics, the motion of particles with a spin of ½ is described as a rule by the Dirac equation with first-order derivatives with respect to space-time variables of the bispinor wavefunction [1]. In [1], Dirac also derived the second-order equations for fermions moving in an external electromagnetic field.

The motion of fermions in external force fields can also be described by self-conjugate equations with spinor wavefunctions [2]. In the transformation of the first-order Dirac equation to the self-conjugate second-order equation, the energy of a spin particle is conserved, while the probability densities for observing particle is different. This leads to new physical consequences. For example, in the presence of event horizons, nonregular solutions to the Dirac equation in the Schwarzschild, Reissner-Nordström, Kerr, and Kerr-Newman external gravitational and electromagnetic fields become regular stationary solutions to second-order equations with square integrable spinor wave functions [3] - [5].

This study is devoted to the analysis of stationary solutions to second-order equations for fermions moving in the Minkowsky space-time in an external Coulomb field.

The article is organized as follows.

In Section 2, for the sake of consistency of presentation, we consider the Dirac equations for a bispinor wavefunction. Non-self-conjugate second-order equations for spinor wavefunctions are also given.

In Section 3, the Dirac equations with spinor wavefunctions are analyzed, and their relationship with the Dirac equations in the Foldy-Wouthuysen representation [6] is established.

In Section 4, self-conjugate second-order equations with spinor wavefunctions are derived for fermions and antifermions in an external Coulomb field. Remarkably, these equations ware derived using closed similarity transformation. After separation of variables, equations for radial wavefunctions are derived. The admissibility of the probabilistic interpretation of spinor wavefunctions, which is identical to the interpretation for wavefunctions of the Klein-Gordon equations with separated states with positive and negative energies, is demonstrated [7].

In Section 5, second-order solutions to the second-order equations in the Coulomb field of attraction are analyzed. It is shown that for identical energy spectra of hydrogen-like atoms, the probability densities of detection of fermions in the corresponding energy states slightly differ from the probability densities calculated using eigenfunctions of the Dirac equation. The difference increases with the growth of the atomic number $Z$.

In Section 6, the second-order equations in the Coulomb field of repulsion are analyzed. The existence of an impenetrable potential barrier with a radius proportional to the classical fermion radius and inversely proportional to the fermion energy is demonstrated (for $E \gg mc^2$), where $E$ and $m$ are the fermion energy and mass and $c$ is the velocity of light. It is shown that the existence of the impenetrable barrier does not contradict the results of experiments on probing of the internal structure of the electron and does not affect (in the lower order of the perturbation theory) the Coulomb electron scattering cross section.

The existence of the impenetrable barrier may lead to confinement of positrons in supercritical nuclei with $Z \gtrsim 170$ during the emission of vacuum electron-positron pairs [8] - [10].

The results obtained in this work are discussed in Conclusions.

## 2. Dirac equation for a bispinor wavefunction

The Dirac equation for the fermions of mass $m$ and charge $e$, moving in an external electromagnetic field, can be written in the form

$$\left[ p^0 - eA^0(\mathbf{r},t) - \boldsymbol{\alpha}\left(\mathbf{p} - e\mathbf{A}(\mathbf{r},t) - \beta m\right) \right] \Psi(\mathbf{r},t) = 0. \tag{1}$$

Here and below, we are using the system of units in which $\hbar = c = 1$ and the following signature of the Minkowsky space-time:

$$g_{\mu\nu} = \text{diag}[1,-1,-1,-1]. \tag{2}$$

In this expression, $\mu, \nu = 0,1,2,3$. In Eq. (1), $\Psi(\mathbf{r},t)$ is the bispinor wavefunction; $A^0(\mathbf{r},t)$ and $\mathbf{A}(\mathbf{r},t)$ are the electromagnetic field potentials; $\alpha^k, \beta$ are four-dimensional Dirac matrices, $k = 1,2,3$; $p^0 = i\dfrac{\partial}{\partial t}$ and $\mathbf{p} = -i\nabla$.

Dirac also derived second-order equation [1]

$$\left[ (p^0 - eA^0)^2 - (\mathbf{p} - e\mathbf{A})^2 - m^2 + e\boldsymbol{\Sigma}\mathbf{H} - i\boldsymbol{\alpha}\mathbf{E} \right] \Psi(\mathbf{r},t) = 0, \tag{3}$$

where $\Sigma = \begin{pmatrix} \boldsymbol{\sigma} & 0 \\ 0 & \boldsymbol{\sigma} \end{pmatrix}$, $\sigma^k$ are two-dimensional Pauli matrices; and $\mathbf{H} = \text{rot}\,\mathbf{A}$, and $\mathbf{E} = -\dfrac{\partial \mathbf{A}}{\partial t} - \nabla A^0$ are the magnetic and electric fields.

Let us suppose that

$$\Psi(\mathbf{r},t) = \begin{pmatrix} \varphi(\mathbf{r},t) \\ \chi(\mathbf{r},t) \end{pmatrix}, \tag{4}$$

where $\varphi(\mathbf{r},t)$ and $\chi(\mathbf{r},t)$ are the spinor wavefunctions.

Then the probability density for Eqs. (1) and (3) is defined as

$$w_D = \Psi^+(\mathbf{r},t)\Psi(\mathbf{r},t) = \varphi^+(\mathbf{r},t)\varphi(\mathbf{r},t) + \chi^+(\mathbf{r},t)\chi(\mathbf{r},t). \tag{5}$$

Equation (1) leads to the following expressions:

$$\begin{aligned}(p^0 - eA^0 - m)\varphi &= \boldsymbol{\sigma}(\mathbf{p} - e\mathbf{A})\chi, \\ (p^0 - eA^0 + m)\chi &= \boldsymbol{\sigma}(\mathbf{p} - e\mathbf{A})\varphi,\end{aligned} \tag{6}$$

$$\begin{aligned}\chi &= \frac{1}{p^0 - eA^0 + m}\boldsymbol{\sigma}(\mathbf{p} - e\mathbf{A})\varphi, \\ \varphi &= \frac{1}{p^0 - eA^0 - m}\boldsymbol{\sigma}(\mathbf{p} - e\mathbf{A})\chi.\end{aligned} \tag{7}$$

### 3. Dirac equations with spinor wavefunctions

Using expressions (7), we obtain from Eq. (3) two separate equations for spinors $\varphi(\mathbf{r},t)$ and $\chi(\mathbf{r},t)$:

$$\left[(p^0 - eA^0)^2 - (\mathbf{p} - e\mathbf{A}) - m^2 + e\boldsymbol{\sigma}\mathbf{H} - i\boldsymbol{\sigma}\mathbf{E}\frac{1}{p^0 - eA^0 + m}\boldsymbol{\sigma}(\mathbf{p} - e\mathbf{A})\right]\varphi(\mathbf{r},t) = 0, \tag{8}$$

$$\left[(p^0 - eA^0)^2 - (\mathbf{p} - e\mathbf{A}) - m^2 + e\boldsymbol{\sigma}\mathbf{H} - i\boldsymbol{\sigma}\mathbf{E}\frac{1}{p^0 - eA^0 - m}\boldsymbol{\sigma}(\mathbf{p} - e\mathbf{A})\right]\chi(\mathbf{r},t) = 0. \tag{9}$$

These equations are non-self-conjugate because the factors in the last terms generally do not commute with each other.

Relations (6) and (7) can also be used for deriving non-self-conjugate equations:

$$\left[p^0 - eA^0 - m - \boldsymbol{\sigma}(\mathbf{p} - e\mathbf{A})\frac{1}{p^0 - eA^0 + m}\boldsymbol{\sigma}(\mathbf{p} - e\mathbf{A})\right]\varphi = 0, \tag{10}$$

$$\left[p^0 - eA^0 + m - \boldsymbol{\sigma}(\mathbf{p} - e\mathbf{A})\frac{1}{p^0 - eA^0 - m}\boldsymbol{\sigma}(\mathbf{p} - e\mathbf{A})\right]\chi = 0. \tag{11}$$

The equations (8) and (10) differ from Eqs. (9) and (11), respectively, in the substitution $p^0 \to -p^0$, $\mathbf{p} \to -\mathbf{p}$, $e \to -e$.

For stationary states, the solutions to Eqs. (8), (10) correspond to states with a positive energy a fermion, while solutions to Eqs. (9), (11) correspond to states with a negative energy:

$$\begin{aligned}p^0\varphi &= |E|\varphi, \\ p^0\chi &= -|E|\chi,\end{aligned} \tag{12}$$

where $E$ is the fermion energy. In the case of stationary states, electromagnetic potentials $A^0(\mathbf{r}), A^k(\mathbf{r})$ are independent of time.

In Eqs. (8), (9) and (10), (11), we separated states with positive and negative energies. In relations (12), the equation for $\varphi$ corresponds to particles, while the equation for $\chi$ corresponds to antiparticles.

Equations (8), (9) and (10), (11) are exact equations with spinor wavefunctions $\varphi(\mathbf{r},t)$ and $\chi(\mathbf{r},t)$. However, these equations are non-self-conjugate. The reason for non-self-conjugation is a transition from the Dirac equation with a spinor wavefunction to Eqs. (8), (9) and (10), (11) with separated states in accordance with the sign of the fermion energy and with spinor wavefunctions.

In quantum mechanics, the Hamiltonians of stationary states are Hermitian and self-conjugate operators. In the absence of the Hamiltonian approach, the operators in the equations for wavefunctions of stationary states must also be Hermitian and self-conjugate. To this end, the equations for wavefunctions must be reduced to the self-conjugate form using nonunitary similarity transformation.

Earlier, an analogous problem of the emerging non-self-conjugation of the equations in a transition from bispinor wavefunctions of the Dirac equation to spinor wavefunctions was also solved using nonunitary similarity transformations (this problem appears in the analysis of relativistic corrections to equation of motion of particles with a spin of ½ in an electrostatic field to within terms on the order of $v^2/c^2$, where $v$ is the particle velocity (see, for example, [11])).

The similarity transformations preserve the energies of states, but do not preserve their probability. In contrast to Eqs. (8) and (9), the advantage of Eqs. (10), (11) is the existence of closed similarity transformations (31) for them.

In deriving Eqs. (8), (9) and (10), (11), the problem of correctness of equalities (7) when their denominators tending to zero can arise. However, in this case (see Section 6) for physically admissible stationary solutions to Eqs. (10), (11), the following asymptotic forms exist:

$$E - eA^0 + m \big|_{r \to r_{cl}} \sim (r - r_{cl}),$$

$$\varphi \big|_{r \to r_{cl}} \sim (r - r_{cl})^2,$$

$$\chi \big|_{r \to r_{cl}} \sim \text{const}\,1,$$

for particles $(E > 0)$ and

$$-|E| + eA^0 - m \big|_{r \to r_{cl}} \sim (r - r_{cl}),$$

$$\chi \big|_{r \to r_{cl}} \sim (r - r_{cl})^2,$$

$$\varphi \big|_{r \to r_{cl}} \sim \text{const}\,2.$$

for antiparticles $(E<0)$. These asymptotics forms (with notation $\mathbf{p}=-i\nabla$) are in conformity with expressions (7). In should be noted that apparent singularity in relations (7) appears only in Coulomb fields of repulsion.

Let us now consider the probabilistic interpretation of wavefunctions in Eqs. (10), (11) in the presence of stationary states (12). Let us begin with Eq. (10):

$$E\varphi = \left(m + eA^0 + \boldsymbol{\sigma}(\mathbf{p}-e\mathbf{A})\right)\frac{1}{E-eA^0+m}\boldsymbol{\sigma}(\mathbf{p}-e\mathbf{A})\varphi. \qquad (13)$$

1. If $A^0(\mathbf{r})=0$, using the method of successive approximations and substituting relations

$$E^0 = m,\ E^{(1)} = m + \frac{[\boldsymbol{\sigma}(\mathbf{p}-e\mathbf{A})]^2}{2m},\ E^{(2)} = m + \frac{[\boldsymbol{\sigma}(\mathbf{p}-e\mathbf{A})]^2}{2m} - \frac{[\boldsymbol{\sigma}(\mathbf{p}-e\mathbf{A})]^4}{8m^3},$$

into the denominator on the right hand side of expressions (13), we can obtain the following closed expression for the Hamiltonian of Eq. (13):

$$E\varphi = H_{FW}^{(+)}\varphi = \left(m + \frac{[\boldsymbol{\sigma}(\mathbf{p}-e\mathbf{A})]^2}{2m} - \frac{[\boldsymbol{\sigma}(\mathbf{p}-e\mathbf{A})]^4}{8m^3} + ...\right)\varphi = \sqrt{m^2 + (\mathbf{p}-e\mathbf{A})^2 - e\boldsymbol{\sigma}\mathbf{H}}\ \varphi. \qquad (14)$$

For stationary states of Eq. (11), an analogous procedure leads to the equation with the minus sign of the square root:

$$E\chi = H_{FW}^{(-)}\chi = -\sqrt{m^2 + (\mathbf{p}-e\mathbf{A})^2 - e\boldsymbol{\sigma}\mathbf{H}}\ \chi. \qquad (15)$$

In expressions (14) and (15), $\mathbf{H}=\operatorname{rot}\mathbf{A}$ is the magnetic field. Equations (14) and (15) show that wavefunctions $\varphi(\mathbf{r},t)$ and $\chi(\mathbf{r},t)$ are the wavefunctions of the Dirac equation in the Foldy-Wouthuysen representation [12]. In this representation, the bispinor wavefunction and the Hamiltonian can be written in form [12], [13]

$$\Psi_{FW}(\mathbf{r},t) = \begin{pmatrix}\varphi(\mathbf{r})\\0\end{pmatrix}e^{-iEt},\ E>0,$$

$$\Psi_{FW}(\mathbf{r},t) = \begin{pmatrix}0\\\chi(\mathbf{r})\end{pmatrix}e^{-iEt},\ E<0, \qquad (16)$$

$$H_{FW} = \beta\sqrt{m^2 + (\mathbf{p}-e\mathbf{A})^2 - e\boldsymbol{\Sigma}\mathbf{H}}. \qquad (17)$$

The expressions hold for any magnetic field magnitude.

2. $A^k(\mathbf{r})=0,\ A^0(\mathbf{r})\neq 0$. In this case, substituting $E^0 = m$ into the denominator of the right-hand side of Eq. (13), we get

$$E^{(1)}\varphi = \left(eA^0 + m + \frac{\mathbf{p}^2}{2m} - \frac{ie\nabla A^0 \mathbf{p}}{4m^2} + \frac{\boldsymbol{\sigma}(e\nabla A^0 \times \mathbf{p})}{4m^2}\right)\varphi. \qquad (18)$$

This equation is non-self-conjugate. To reduce it to the self-conjugate form, we must perform nonunitary similarity transformation

$$\varphi_{sc}^{(1)} = g_{\varphi}^{(1)} \varphi, \qquad (19)$$

where we can write to within $\mathbf{p}^2$ and $eA^0$,

$$g_{\varphi}^{(1)} = 1 + \frac{\mathbf{p}^2}{8m^2} + \frac{eA^0}{2m},$$

$$\left(g_{\varphi}^{(1)}\right)^{-1} = 1 - \frac{\mathbf{p}^2}{8m^2} - \frac{eA^0}{2m}. \qquad (20)$$

This gives

$$E^{(1)}\varphi_{sc}^{(1)} = H_{FW}^{(1)}\varphi_{sc}^{(1)} = g_{\varphi}^{(1)}\left(eA^0 + m + \frac{\mathbf{p}^2}{2m} - \frac{ie\nabla A^0 \mathbf{p}}{4m^2} + \frac{\boldsymbol{\sigma}(e\nabla A^0 \times \mathbf{p})}{4m^2}\right)\left(g_{\varphi}^{(1)}\right)^{-1}\varphi_{sc}^{(1)} =$$

$$= \left(eA^0 + m + \frac{\mathbf{p}^2}{2m} + \frac{\Delta eA^0}{8m^2} + \frac{\boldsymbol{\sigma}(e\nabla A^0 \times \mathbf{p})}{4m^2}\right)\varphi_{sc}^{(1)}. \qquad (21)$$

where $\Delta = \nabla^2$ is the Laplacian.

Equation (11) for states with $E < 0$ has the form

$$E\chi = \left(eA^0 - m + \boldsymbol{\sigma}\mathbf{p}\frac{1}{E - eA^0 - m}\boldsymbol{\sigma}\mathbf{p}\right)\chi. \qquad (22)$$

Substituting $E^{(0)} = -m$ into the denominator on the right-hand side of Eq. (22), we obtain

$$E^{(1)}\chi = \left(eA^0 - m - \frac{\mathbf{p}^2}{2m} - \frac{ie\nabla A^0 \mathbf{p}}{4m^2} + \frac{\boldsymbol{\sigma}(e\nabla A^0 \times \mathbf{p})}{4m^2}\right)\chi. \qquad (23)$$

To deduce this equation to the self-conjugate form, we can use equality

$$\chi_{sc} = g_{\chi}^{(1)}\chi, \qquad (24)$$

where we can write correct to $\mathbf{p}^2$ and $eA^0$,

$$g_{\chi}^{(1)} = 1 + \frac{\mathbf{p}^2}{8m^2} - \frac{eA^0}{2m},$$

$$\left(g_{\chi}^{(1)}\right)^{-1} = 1 - \frac{\mathbf{p}^2}{8m^2} + \frac{eA^0}{2m}. \qquad (25)$$

This gives

$$E^{(1)}\chi_{sc}^{(1)} = \left(eA^0 - m - \frac{\mathbf{p}^2}{2m} + \frac{\Delta eA^0}{8m^2} + \frac{\boldsymbol{\sigma}(e\nabla A^0 \times \mathbf{p})}{4m^2}\right)\chi_{sc}. \qquad (26)$$

Equations (21) and (26) are the first terms of the expansion of the Foldy-Wouthuysen Hamiltonian in the external field $A^0(\mathbf{r})$ [6].

Performing the successive approximation procedure for stationary solutions to Eqs. (10) and (11), we can obtain the Foldy-Wouthuysen Hamiltonian in the form of a power series in $\mathbf{p}^2$ and $eA^0$. At each stage of successive approximations, the wavefunctions of Eqs. (10) and (11) are subjected to the corresponding nonunitary similarity transformations:

$$\Psi_{FW}(\mathbf{r},t) = \begin{pmatrix} \varphi^{(n)}(\mathbf{r}) \\ 0 \end{pmatrix}\bigg|_{n\to\infty} e^{-iEt}, \quad E > 0,$$

$$\Psi_{FW}(\mathbf{r},t) = \begin{pmatrix} 0 \\ \chi^{(n)}(\mathbf{r}) \end{pmatrix}\bigg|_{n\to\infty} e^{-iEt}, \quad E < 0,$$

(27)

where

$$\varphi^{(n)}(\mathbf{r})\big|_{n\to\infty} = g_\varphi^{(n)} \varphi(\mathbf{r}),$$
$$\chi^{(n)}(\mathbf{r})\big|_{n\to\infty} = g_\chi^{(n)} \chi(\mathbf{r}).$$

(28)

In the presence of stationary states, analysis of Eqs. (10) and (11) with spinor wavefunctions leads to the following conclusions.

1. For $A^0(\mathbf{r}) = 0$ and $A^k(\mathbf{r}) \neq 0$, Eqs. (10) and (11) can be reduced to the Dirac equation in the Foldy-Wouthuysen representation with separated states having positive and negative fermion energies. Hamiltonian $H_{FW}$ has the closed form (see expression (17)).

2. For $A^0(\mathbf{r}) \neq 0$ and $A^k(\mathbf{r}) = 0$, Hamiltonian $H_{FW}$ can be obtained only in form of a power series in $\mathbf{p}^2$ and $e\mathbf{A}^0$ (first terms of the series in (21) and (26)).

3. For $A^0(\mathbf{r}) \neq 0$, a transformation to the Foldy-Wouthuysen representation with Hermitian Hamiltonian $H_{FW}$ is performed by reducing Eqs. (10) and (11) to the self-conjugate form. In this case, the equations and wavefunctions $\varphi(\mathbf{r},t)$ and $\chi(\mathbf{r},t)$ are subjected to nonunitary similarity transformation.

4. In the Foldy-Wouthuysen representation, the probability density is given, in accordance with relations (27), by

$$w_\varphi = \varphi^{(n)+}\varphi^{(n)}\big|_{n\to\infty}, \quad E > 0,$$
$$w_\chi = \chi^{(n)+}\chi^{(n)}\big|_{n\to\infty}, \quad E < 0.$$

(29)

Since the transition to the Foldy-Wouthuysen representation can be performed via unitary transformation of the Dirca equation with a bispinor wavefunction [14], [15], probability densities (29) coincide with Dirac probability density (5).

# 4. Self-conjugate second-order equations with spinor wavefunctions and probabilistic interpretation

By premultiplying Eq. (10) by operator $\left(p^0 - eA^0 + m\right)$ and Eq. (11) by operator $\left(p^0 - eA^0 - m\right)$, we obtain Eqs. (8) and (9). For stationary states (12), the solutions to Eqs. (8) and (9), as well as Eqs. (10) and (11) are separated state with positive and negative fermion energies.

For $A^0(\mathbf{r}) = 0$ in the present of stationary states, Eqs. (8) and (9) are self-conjugate; their solutions coincide with the solutions to the Dirac equations in the Foldy-Wouthuysen representation (see relations (14), (15)).

Let us suppose that $A^0(\mathbf{r}) \neq 0$ and $A^k(\mathbf{r}) = 0$. In this case, Eqs. (8) and (9) with substitution $p^0 \to E$ (in the presence of stationary states) must be reduced to the self-conjugate form. Remarkably, in contrast to Eqs. (10), (11), the transformation operators for Eqs. (8), (9) can be written in closed form,

$$\Phi = g_\varphi \varphi,$$
$$\mathrm{X} = g_\chi \chi, \tag{30}$$

where

$$g_\varphi = \left(E - eA^0 + m\right)^{-1/2},$$
$$g_\chi = \left(|E| + eA^0 + m\right)^{-1/2}. \tag{31}$$

After transformations, Eqs. (8) and (9) have the self-conjugate form

$$g_\varphi \left[\left(E - eA^0\right)^2 - \mathbf{p}^2 - m^2 - \frac{1}{E - eA^0 + m} i e \boldsymbol{\sigma} \mathbf{E} \boldsymbol{\sigma} \mathbf{p}\right] g_\varphi^{-1} \Phi =$$
$$= \left[\left(E - eA^0\right)^2 - \mathbf{p}^2 - m^2 - \frac{3}{4} \frac{e^2 \mathbf{E}^2}{\left(E - eA^0 + m\right)^2} + \frac{e \, \mathrm{div}\mathbf{E}}{2\left(E - eA^0 + m\right)} + \frac{e\boldsymbol{\sigma}(\mathbf{E} \times \mathbf{p})}{E - eA^0 + m}\right] \Phi = 0, \tag{32}$$

$$g_\chi \left[\left(E - eA^0\right)^2 - \mathbf{p}^2 - m^2 - \frac{1}{E - eA^0 - m} i e \boldsymbol{\sigma} \mathbf{E} \boldsymbol{\sigma} \mathbf{p}\right] g_\chi^{-1} \mathrm{X} =$$
$$= \left[\left(E - eA^0\right)^2 - \mathbf{p}^2 - m^2 - \frac{3}{4} \frac{e^2 \mathbf{E}^2}{\left(E - eA^0 - m\right)^2} + \frac{e \, \mathrm{div}\mathbf{E}}{2\left(E - eA^0 - m\right)} + \frac{e\boldsymbol{\sigma}(\mathbf{E} \times \mathbf{p})}{E - eA^0 - m}\right] \mathrm{X} = 0. \tag{33}$$

Equation (32) describes particles with a spin of ½ and with a positive energy. Equation (33) describes antiparticles with negative energies. Equations (32) and (33) differ in the sign of the mass term. Both the equations can be written in the form typical of the Foldy-Wouthuysen representation:

$$\left[\left(E-eA^0\right)^2-\mathbf{p}^2-m^2-\frac{3}{4}\frac{e^2\mathbf{E}^2}{\left(E-eA^0+\beta m\right)^2}+\frac{e\,\text{div}\mathbf{E}}{2\left(E-eA^0+\beta m\right)}+\frac{e\boldsymbol{\sigma}\left(\mathbf{E}\times\mathbf{p}\right)}{E-eA^0+\beta m}\right]\Psi(\mathbf{r},t)=0, \quad (34)$$

where $\Psi(\mathbf{r},t)$ is a bispinor,

$$\Psi(\mathbf{r},t)=\begin{pmatrix}\Phi(\mathbf{r})\\0\end{pmatrix}e^{-iEt},\ E>0,$$
$$\Psi(\mathbf{r},t)=\begin{pmatrix}0\\X(\mathbf{r})\end{pmatrix}e^{-iEt},\ E<0. \quad (35)$$

It is well known (see, for example, [7]), that the probabilistic interpretation is admissible for separated states with positive and negative energy for the Klein-Gordon equation. The probability densities for Eq. (34) and Eqs. (32), (33) are given by

$$w_\Phi=\Phi^+\Phi,\ E>0,$$
$$w_X=X^+X,\ E<0. \quad (36)$$

Since transformations (31) differ from transformations (20), (25) and (28), probability densities (36) differ from probability densities (5) and (29) for the Dirac equation. The extent of these differences will be considered in the next section of this article.

Similarity transformations preserve the energies of states. The energies of a particle with a spin of ½ in second-order equations (32), (33) coincide with corresponding energies in the Dirac equation (1) in the Foldy-Wouthuysen representation.

For a centrally symmetric Coulomb potential, Eqs. (32), (33) permit separation of variables in spherical coordinates $(r,\theta,\varphi)$.

If we write spinors $\Phi(\mathbf{r})$ and $X(\mathbf{r})$ in form

$$\Phi(r,\theta,\varphi)=F(r)\Omega_{jlm_\varphi}(\theta,\varphi), \quad (37)$$

$$X(r,\theta,\varphi)=G(r)(-1)^{\frac{1+l-l'}{2}}\Omega_{jl'm_\varphi}(\theta,\varphi), \quad (38)$$

where $\Omega_{jlm_\varphi},\Omega_{jl'm_\varphi}$ are spherical spinors [16]; $l=j\pm(1/2)$, $l'=2j-1$; $j,l$ being the total and orbital angular momenta of a particle with a spin of ½; $m_\varphi=-j,-j+1,...,j$ is the projection of total angular momentum $j$, we can obtain the Schrödinger-type equation with effective potentials $U^F_{eff}(r),U^G_{eff}(r)$ for radial functions $F(r)$ and $G(r)$:

$$\frac{1}{r^2}\frac{d}{dr}\left(r^2\frac{d}{dr}\right)F(r)+2\left(E_{Schr}-U^F_{eff}(r)\right)F(r)=0, \quad (39)$$

$$\frac{1}{r^2}\frac{d}{dr}\left(r^2\frac{d}{dr}\right)G(r)+2\left(E_{Schr}-U^G_{eff}(r)\right)G(r)=0. \quad (40)$$

In these equations,

$$E_{Schr} = (E^2 - m^2)/2. \tag{41}$$

$$U_{eff}^F(r) = EeA^0 - \frac{1}{2}(eA^0)^2 + \frac{\kappa(\kappa+1)}{2r^2} + \frac{1}{4}\frac{\dfrac{d^2(eA^0)}{dr^2}}{E+m-eA^0} + \frac{3}{8}\frac{\left(\dfrac{deA^0}{dr}\right)^2}{(E+m-eA^0)^2} - \frac{1}{2}\frac{\kappa \dfrac{deA^0}{dr}}{r(E+m-eA^0)}. \tag{42}$$

$$U_{eff}^G(r) = EeA^0 - \frac{1}{2}(eA^0)^2 + \frac{\kappa(\kappa+1)}{2r^2} + \frac{1}{4}\frac{\dfrac{d^2(eA^0)}{dr^2}}{E-m-eA^0} + \frac{3}{8}\frac{\left(\dfrac{deA^0}{dr}\right)^2}{(E-m-eA^0)^2} - \frac{1}{2}\frac{\kappa \dfrac{deA^0}{dr}}{r(E-m-eA^0)}. \tag{43}$$

In Eqs. (42), (43), $\kappa$ is the quantum number of the Dirac equation,

$$\kappa = \mp 1, \mp 2, \ldots = \begin{cases} -(l+1), & j = l+1/2, \\ l, & j = l-1/2. \end{cases} \tag{44}$$

Below, we will use Eq. (39) for analyzing the motion of fermions in Coulomb fields of various intensities.

## 5. Second-order equations in a Coulomb field of attraction

Let us consider solutions to Schrödinger-type equation (39) for fermions in Coulomb field $eA^0(r) = -Ze^2/r$, where $Z$ is the atomic number. We will use below dimensionless quantities $\varepsilon = E/m$ and $\rho = r/l_c$, where $l_c = \hbar/mc$ is the fermion Compton wavelength.

The asymptotic form of effective potential (42) for $r \to 0$ has the form

$$U_{eff}^F\big|_{\rho \to 0} = -\frac{(Z\alpha)^2 - (3/4) + (1-\kappa^2)}{2\rho^2}. \tag{45}$$

In (45) $\alpha = e^2/\hbar c$ is the fine-structure electromagnetic constant.

Depending on $Z$, we can single out three characteristic regions in asymptotic form (45). By way of example, we consider these regions for bound states $1S_{1/2}$ ($\kappa = -1$) and $2P_{1/2}$ ($\kappa = +1$). In the first region $1 \leq Z < \sqrt{3}/2\alpha$ there exists positive barrier a $\sim 1/\rho^2$ followed by a potential well. For $Z = Z_{cr} = \sqrt{3}/2\alpha \approx 118.7$, the potential barrier disappears; for $Z > Z_{cr}$ for $\rho \to 0$, the potential well $-K/\rho^2$ remains. In the second region $119 \leq Z \leq 137$, there is the potential well with $K \leq 1/8$, which permits the existence of fermion stationary bound states [17]. In the third region $Z > 137$ for $\rho \to 0$, a potential well with $K > 1/8$ exists, which indicates the realization of the regime of "falling to the center" [17]. Dependences $U_{eff}^F(\rho)$ for $\kappa = -1$ ($1S_{1/2}$) are shown in Fig. 1 for $Z = 1, 119$ and $140$.

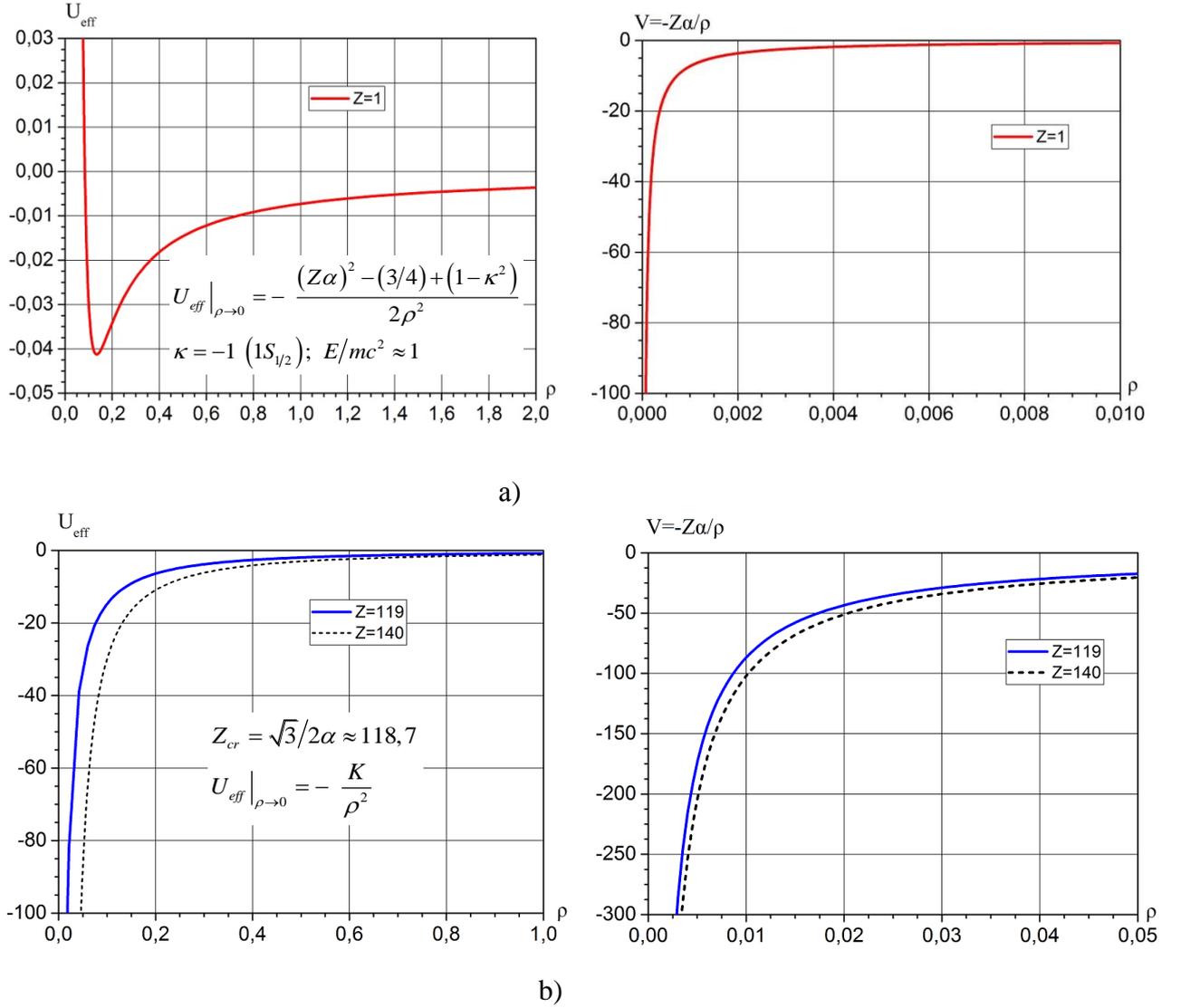

Fig.1. Dependences $U_{eff}(\rho)$ and $V(\rho)$ for (a) $Z=1$, (b) 119, 140.

For comparison, the dependencies (for $Z=1, 119$ and $140$) of the Coulomb potential $eA^0(\rho) = -Z\alpha^2/r$, which is used in the Dirac equations, are also shown in Fig. 1. In can be seen that dependencies $U_{eff}^F(\rho)$ and $eA^0(\rho)$ differ significantly. It should be noted that the above three regions can also be obtained from rigorous mathematical analysis of the Dirac equation with the singular potential $eA^0(\rho)$ [18], [19]. In should also be noted that the problem of "falling to the center" for $Z > 137$ can be solved considering finite sizes of atomic nuclei [8] - [10]. Analogous analysis of the asymptotic forms of $U_{eff}^F(\rho)$ (42) with another division into three regions can be performed for admissible values of $\kappa \neq \pm 1$.

We have determined the energy spectrum and eigenfunctions of Schrödinger-type equations (39) numerically using the Pruefer transformation [20] and the method of phase functions [21] - [23]. Some details are given in Appendix. This method of numerical solution of Schrodinger-type equations applied for describing the motion of fermions in the Schwarzschild, Reissner-Nordström and Kerr-Newman fields was described in detail in [3] - [5].

It was shown as a result of calculations that the energy spectrum of second-order equation (39), as expected, coincides with the spectrum of hydrogen-like atoms, which can be obtained by solving Dirac equation (1).

A slightly different situation appears when probability densities (5) (with eigenfunctions of Dirac equation (1)) are compared with probability densities (36) (with eigenfunctions of Eqs. (32), (36)). Our further analysis will be confined to comparison of probability densities with radial wavefunctions of Dirac equation (1) and Eq. (32) for states with $E > 0$. Corresponding analysis using Eq. (33) for states with $E < 0$ can be performed analogously.

In the representation of the bispinor wavefunctions, we can single out radial functions $\varphi(\rho)$ and $\chi(\rho)$:

$$\varphi(\mathbf{\rho},t) = \varphi(\rho)\Omega_{jlm_\varphi}(\theta,\varphi)e^{-i\varepsilon t},$$
$$\chi(\mathbf{\rho},t) = \chi(\rho)(-1)^{\frac{1+l-l'}{2}}\Omega_{jl'm_\varphi}(\theta,\varphi)e^{-i\varepsilon t}, \quad (46)$$

where $\Omega_{jlm_\varphi}, \Omega_{jl'm_\varphi}$ are spherical spinors [16] (see also expressions (37) and (38)).

Then, in accordance with relation (5), the probability of detecting a fermion at distance $\rho$ in spherical layer $d\rho$ for Dirac wavefunctions is given by

$$w_D(\rho) = \left(\varphi^*(\rho)\varphi(\rho) + \chi^*(\rho)\chi(\rho)\right)\rho^2 d\rho. \quad (47)$$

An analogous probability with the eigenfunctions of second-order equation (39) is (see Eqs. (36), (37), (30))

$$w_\Phi(\rho) = F^*(\rho)F(\rho) = \frac{1}{\varepsilon + (Z\alpha^2/\rho) + 1}\varphi^*(\rho)\varphi(\rho)\rho^2 d\rho. \quad (48)$$

Even at this stage, it can be seen without calculations that probabilities (47) and (48) are different. Coulomb functions $\chi(\rho)$ of hydrogen-like atoms do not vanish in the entire domain $(0,\infty)$, while the functions $\varphi(\rho)$ vanish one or several times for excited energy states [24]. Probability (48) is proportional to $\varphi^*(\rho)\varphi(\rho)$; therefore, it differs from probability (47). Let us consider the extent of these differences. Figure 2 shows probabilities (47), (48) calculated for states of $1S_{1/2}, 2S_{1/2}, 2P_{1/2}, 3P_{1/2}$, and $Z = 1$ and $100$.

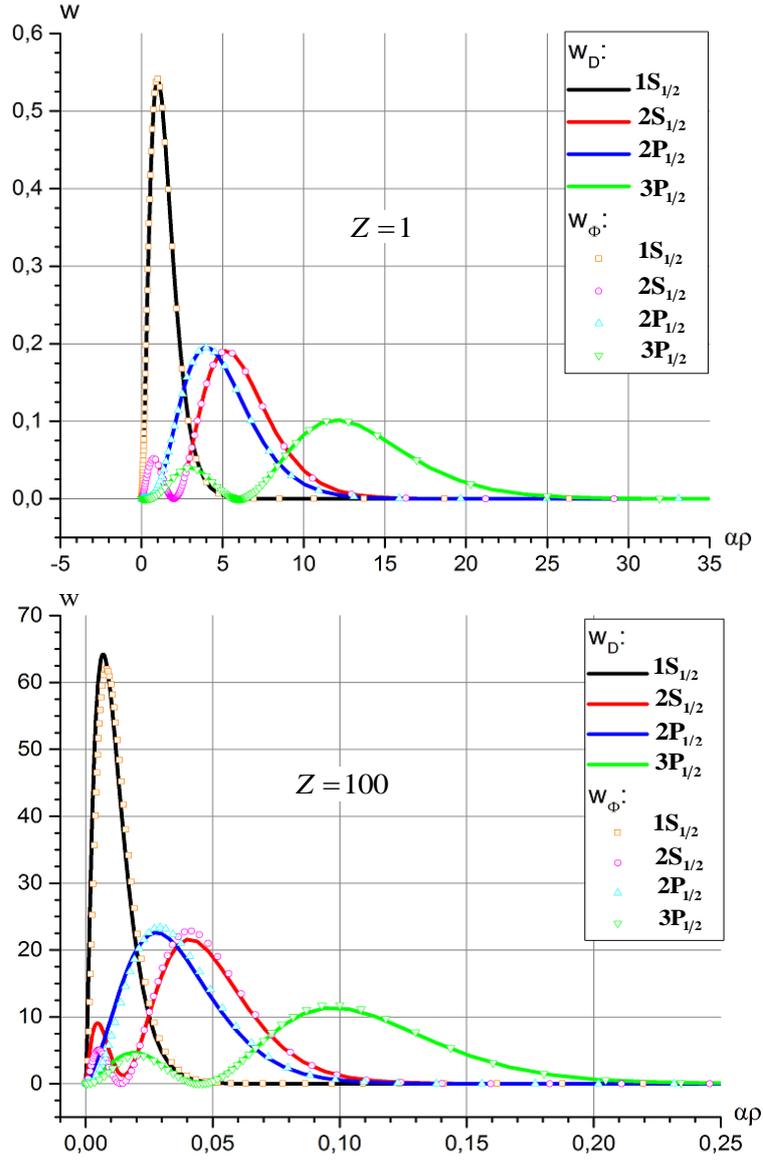

Fig. 2. Comparison of probability densities $w_D(\rho)$ and $w_\Phi(\rho)$ for (a) $Z=1$ and (b) 100.

In Fig. 2, the distances on the abscissa axis are given in atomic units $\rho_a = (r/\hbar^2)me^2 = \alpha\rho$.

For the hydrogen atom $(Z=1)$, the relative contribution of term $\chi^*(\rho)\chi(\rho)$ in expression (47) is on the order of $\sim 10^{-7}$. It can be seen that probability densities (47) and (48) coincide to such a degree of accuracy.

For the hydrogen-like atom with $Z=100$, the weight of term $\chi^*(\rho)\chi(\rho)$ is approximately 16%. For qualitatively coinciding dependences $w_D(\rho)$ and $w_\Phi(\rho)$, slight differences can also be seen. The effect of these differences on the matrix elements of actual physical processes requires additional analysis. It is obvious, however, that this effect is small, and we can conclude on the whole that the second-order equation with spinor wavefunctions and the Dirac equation describe the motion of fermions in a Coulomb field of attraction with identical energy spectra and with close probability densities.

## 6. Second-order equations in the Coulomb field of repulsion. Impenetrable potential barriers

In this case, the Coulomb potential is

$$eA^0(\rho) = +Z\alpha^2/\rho. \tag{49}$$

In contrast to this relation, effective potential (42) is singular at

$$\rho_{cl} = Z\alpha/(\varepsilon+1) \tag{50}$$

In natural units radius $r_{cl}$ is proportional to classical radius $r_f = +Ze^2/mc^2$ of a charged fermion

$$r_{cl} = \frac{r_f}{1+E/mc^2} = \frac{Z(e^2/mc^2)}{1+E/mc^2}. \tag{51}$$

The leading asymptotic form of potential (42) for $\rho \to \rho_{cl}$ has form

$$U^F_{eff}\bigg|_{\rho \to \rho_{cl}} = \frac{3}{8} \frac{1}{(\rho-\rho_{cl})^2}. \tag{52}$$

If we write the radial function $F(\rho)$ for $\rho \to \rho_{cl}$ as

$$F(\rho)\bigg|_{\rho \to \rho_{cl}} = (\rho-\rho_{cl})^s \sum_{k=0}^{\infty} f_k (\rho-\rho_{cl})^k, \tag{53}$$

the indicial equation for Eq. (39) combined with (52) leads to two solutions $s_1 = 3/2, s_2 = -1/2$, and

$$F_1\bigg|_{\rho \to \rho_{cl}} = f_0^{(1)} (\rho-\rho_{cl})^{3/2}, \tag{54}$$

$$F_2\bigg|_{\rho \to \rho_{cl}} = f_0^{(2)} (\rho-\rho_{cl})^{-1/2}. \tag{55}$$

Solution (55) is square nonintegrable and hence, is physically inadmissible.

Solution (54) combined with (30), (31), (37), and (46) leads to the following dependences of Dirac radial functions $\varphi(\rho), \chi(\rho)$:

$$\chi\bigg|_{\rho \to \rho_{cl}} \sim (\rho-\rho_{cl})^2. \tag{56}$$

It was shown earlier numerically for solutions to the Dirac equation in the Kerr-Newman field with $\varepsilon < 1$ for the case with potential barrier (52) followed by a potential well that radial functions behave in exact conformity with asymptotic form (56) [5].

However, the Coulomb functions of the continuous spectrum of hydrogen-like atoms with $\varepsilon > 1$ for potential (49) do not vanish for $\rho = \rho_{cl}$, but equal certain constants [16]. Such a behavior of $\chi(\rho)$ leads to nonphysical asymptotic form (55) of the wavefunction of second-order equation (39). Obviously, the second solution to the Dirac equation with asymptotic form (56) must also exist for the Coulomb potential of repulsion.

Numerical analysis of integral curves of solutions to Dirac equation (1) for radial functions $\varphi(\rho), \chi(\rho)$ shows that the solution with asymptotic form (56) exists indeed. Remarkably, at distances of several Compton wavelength from $\rho = \rho_{cl}$, the solution begins to coincide with the Coulomb functions of the continuous spectrum with $\varepsilon > 1$.

Figure 3 shows dependences $U_{eff}^F(\rho)$ for $Z=1, \kappa=-1, \varepsilon \simeq 1$ and $Z=140, \kappa=-1, \varepsilon \approx 1$. Dependences $eA^0(\rho) = Ze^2/\rho$ for $Z=1, Z=140$ are also shown for comparison.

For asymptotic form (54), potential barrier (52) is quantum-mechanically impenetrable [25]. The following question arises: does the existence of the impenetrable barrier for $\rho = \rho_{cl}$ corresponds to the available experimental data?

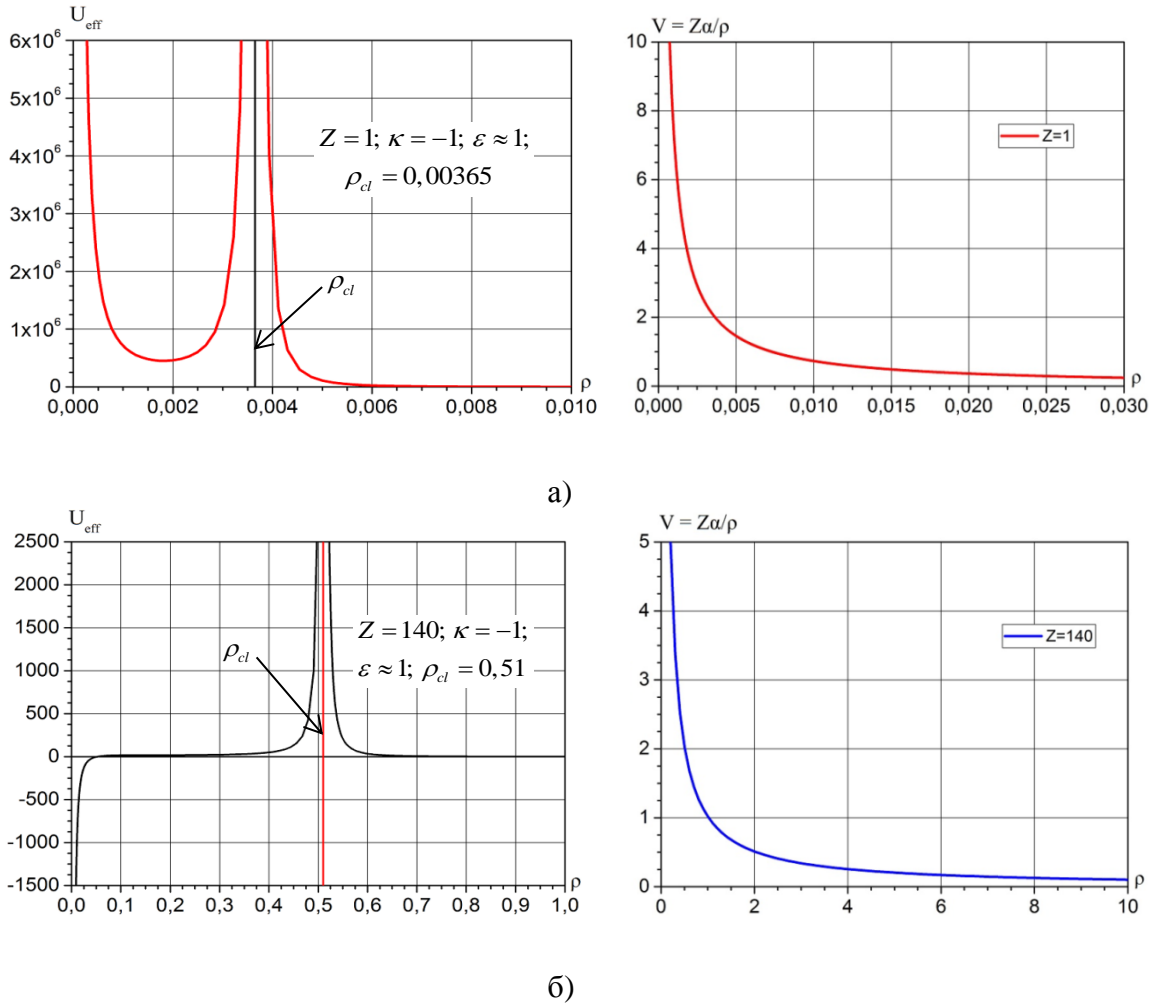

Fig. 3. Dependences $U_{eff}(\rho)$ and $V(\rho)$ for (a) $Z=0$ and (b) 140.

## 6.1 Probing of the electron internal structure

For electron scattering by an electron, reduced mass $m = m_e/2$, where $m_e$ is the electron mass. The expression for impenetrable barrier radius (51) assumes the form

$$r_{cl}^{(e)} = \frac{2r_e}{1 + 2E/m_e c^2}. \qquad (57)$$

In this expression, $r_e = e^2/m_e c^2$ is the classical electron radius. For an electron at rest $(E = m_e c^2)$, the barrier radius is equal to $2/3$ of the classical radius. For $E \gg m_e c^2$, the barrier radius decreases in inverse proportion to the electron energy in the center-of-mass system.

For probing the internal structure of the electron, the impossibility of penetration of electrons to region $r < r_{cl}^{(e)}$ must be fixed in all possible experiments with electron-electron scattering.

In experiments with electron-positron scattering performed at the end of the last century on the LEP accelerator (CERN), for energy $E = 200\text{GeV}$ of electrons and positrons in the center-of-mass system, the internal structure is not manifested for $r_{min} \approx 2 \cdot 10^{-18}$ cm (see, for example, [26]). Obviously, in analogous hypothetical experiments with electron- electron scattering, the value of $r_{min}$ will be slightly larger. Impenetrable barrier radius (57) for $E = 200\text{GeV}$ is $r_{cl}^{(e)} = 0,7 \cdot 10^{-18}$ cm; i.e.,

$$r_{cl}^{(e)} < r_{min}. \qquad (58)$$

At this stage, the existence of impenetrable barrier (52) with radius (57) does not contradict the experimental data on probing the external structure of electron.

## 6.2 Scattering cross section in a Coulomb field

We consider the Coulomb scattering of an electron in the lowest order in $e$ using self-conjugate second-order equation (32) with the spinor wavefunction $\Phi(\mathbf{r})$.

The Feynman diagram has the form

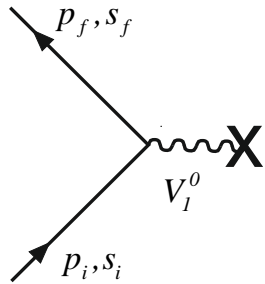

In the first order in $e$, vertex operator $V_1^0$ is given by (see Eq. (32))

$$V_1^0 = e\left[E_f + E_i - \frac{\mathrm{div}\mathbf{E}}{E_i + m} - \frac{\boldsymbol{\sigma}(\mathbf{E}\times\mathbf{p})}{E_i + m}\right]. \tag{59}$$

As compared to the Klein-Gordon equation for spinless particles, the existence of the spin leads to the emergence of the last two terms in vertex operator (59). For the static Coulomb potential, we have

$$A^\mu(q) \equiv \int d^4 x e^{iqx} A^\mu(x) = \frac{Ze}{|\mathbf{q}|^2} 2\pi\delta(E_f - E_i) g^{\mu o}. \tag{60}$$

In this relation, $q = p_f - p_i$; $p_f^0 = p_i^0$; $\mathbf{q} = \mathbf{p_f} - \mathbf{p_i}$; $\mathbf{p_f^2} = \mathbf{p_i^2}$.

Considering relation (60), we obtain

$$\mathbf{E} = -\nabla A^0 = i\mathbf{q} A^0(q); \mathrm{div}\mathbf{E} = \mathbf{q}^2 A^0(q); \boldsymbol{\sigma}(\mathbf{E}\times\mathbf{p_i}) = -i\boldsymbol{\sigma}(\mathbf{q}\times\mathbf{p_i}) A^0(q).$$

Ultimately, vertex operator (59) is given by

$$V_1^0 = e\left(E_i + m + \frac{\boldsymbol{\sigma}\mathbf{p_f}\boldsymbol{\sigma}\mathbf{p_i}}{E_i + m}\right). \tag{61}$$

Free solutions to the Klein-Gordon equation and Eqs. (32), (33) in the form of plane waves in the continuous spectrum can be written in form

$$\Phi = \frac{1}{(2\pi)^{3/2}} \frac{1}{\sqrt{2E_p}} e^{-ipx} \Phi_s,$$
$$X = \frac{1}{(2\pi)^{3/2}} \frac{1}{\sqrt{2E_p}} e^{ipx} X_s. \tag{62}$$

Here, $E_p = \sqrt{m^2 + \mathbf{p}^2}$; $\Phi_s$ and $X_s$ are normalized Pauli spin functions.

As the result, the transition amplitude for electron scattering in a Coulomb field can be written as

$$S_{fi} = -i\int d^4 x \Phi^+(x, p_f, s_f) V_1^0 A^0(x) \Phi(x, p_i, s_i) =$$
$$= -\frac{i\delta(E_f - E_i)}{(2\pi)^2} \Phi_{s_f}^+ \langle\mathbf{p_f}|V_1^0 A^0|\mathbf{p_i}\rangle \Phi_{s_i} = -i\frac{Ze^2}{\mathbf{q}^2} \frac{\delta(E_f - E_i)}{(2\pi)^2} \Phi_{s_f}^+ \frac{1}{2E_i}\left(E_i + m + \frac{1}{E_i + m}\boldsymbol{\sigma}\mathbf{p_f}\boldsymbol{\sigma}\mathbf{p_i}\right)\Phi_{s_i} \tag{63}$$

This amplitude coincides with corresponding amplitude $S_{fi}$ that was calculated earlier in [14] for the Foldy-Wouthuysen representation.

Furthermore, using conventional methods and matrix element $S_{fi}$, we can obtain the Mott differential scattering cross section transformed into the Rutherford cross section in the nonrelativistic case (see, for example, [7]).

Thus, we have proven that in the lowest order of the perturbation theory $(\sim e^4)$, the existence of impenetrable barrier (52) does not affect in the Coulomb electron scattering cross section. To draw conclusions concerning the magnitudes of radiation corrections to the scattering cross section, it is necessary to calculate higher-order diagrams in the perturbation theory.

### 6.3 Positron confinement in supercritical nuclei

An impenetrable barrier (52) for positrons exists in the case of Coulomb field $eA^o(r) = -Ze^2/r$. For completely ionized stable atoms, the range of $Z$ variations in expresion (51) is from $Z = 1$ (hydrogen atom nucleus) to $Z = 92$ (uranium $^{238}U$ nucleus). In accordance with expression (51), the radius of barrier (52) for a positron at rest in the center-of-mass system varies from $1.4 \cdot 10^{-13}$ cm $(Z=1)$ to $1.3 \cdot 10^{-11}$ cm $(Z=92)$. Upon a further hypothetic increase of $Z$ to $Z = Z_{scr} \approx 170$, the energy level of the electronic ground state in the Coulomb field of the nucleus reaches the upper boundary of lower continuum $E = -m_e c^2$. Then for the $S$ states in the $K$-shell unoccupied by electrons, two positron-electron pairs are produced [8] - [10]. Electrons occupy empty $K$ shell, the atomic charge decreases by two units, and positrons leave the atomic system where they can be detected in principle.

However, in the presence of impenetrable barrier (52), the variant of positron confinement is possible. Generated positrons are inside the impenetrable barrier with $r = r_{cl}$ (see Fig.4). Subsequently, the positrons annihilate with electrons of the $K$ shell with the emission of $\gamma$ quanta, and the system returns to its initial state with $Z = Z_{scr}$, and the events can be repeated in the future. The source of information on processes in an atomic system with $Z = Z_{scr}$ for external world is the source of $\gamma$ quanta resulting from annihilation of electrons with positrons.

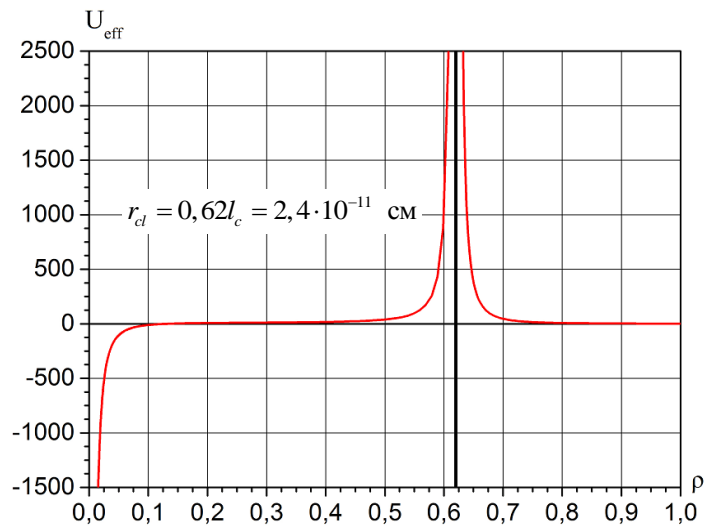

Fig. 4. Dependence $U_{eff}(\rho)$ for a positron moving in a Coulomb field of repulsion with $Z_{scr} = 170$.

## 7. Conclusions

For a quantum-mechanical description of motion of fermion in an external Coulomb field, we have derived self-conjugate second-order equations with spinor wavefunctions. For stationary states, the equations are characterized by separated states with positive and negative energies, which make probabilistic interpretation possible. For a Coulomb field of attraction, the energy spectrum of the second-order equation coincides with the spectrum of the Dirac equation. The probability densities for the second-order equation slightly differ from the probability densities of states for Dirac equation. This difference increases with atomic number $Z$.

When the second-order equation is used for the Coulomb field of repulsion, there exists an impenetrable barrier with a radius depending on the classical electron radius and on the electron energy:

$$r_{cl} = \frac{Ze^2}{mc^2} \frac{1}{1+\left(E/mc^2\right)}.$$

The existence of the impermeable barrier does not contradict experimental data on probing of the electron internal structure and does not affect the Coulomb electron scattering cross section in the lowest order of perturbation theory $\left(\sim e^4\right)$.

The presence of the impenetrable barrier can lead to confinement of positrons in supercritical nuclei with $Z \gtrsim 170$ in the case of spontaneous emission of electron-positron pairs by vacuum.

The existence of the impenetrable barrier indirectly reveals the relation between the spin and charge of elementary particles. Indeed, for spinless particles and unlike charges of the spin particle and the Coulomb potential, the impenetrable barrier does not exist, while for like charges of the spin particle and the Coulomb potential, the impenetrable barrier is present. The location of the barrier charges with the particle energy. Future models of the internal structure of leptons must describe qualitatively the qualitative relationship between the spin and charge of particles, which has been established above.

To use of self-conjugate second-order equations with spinor wavefunctions has already led to new physical results.

For external Schwarzschild, Reissner-Nordström, Kerr, and Kerr-Newman gravitational and electromagnetic fields, we have obtained regular stationary solutions to the second-order equations with square integrable spinor wavefunctions [3] - [5]. These solutions for the Dirac equation are nonregular and square nonintegrable.

Our analysis has shown that the use of second-order self-conjugate equations with spinor wavefunctions extends the possibilities of obtaining regular solutions of quantum mechanics in external gravitational and electromagnetic fields. The solutions to the second-order equations in the Minkowsky space-time do not contradict the available experimental data at the given stage of investigations.

# APPENDIX

## Pruefer transformation and boundary conditions

For numerical solution of Eqs. (39) and (40), we have used the Pruefer transformation [20] - [23].

By way of example, let us consider Eq. (39).

For function $f(r) = rF(r)$, Eq. (39) has form

$$\frac{d^2 f(r)}{dr^2} + 2\left(E_{Schr} - U_{eff}^F(r)\right) f(r) = 0. \tag{A.1}$$

Let us assume that

$$f(r) = P(r)\sin\Phi(r),$$
$$\frac{df(r)}{dr} = P(r)\cos\Phi(r). \tag{A.2}$$

Then

$$\frac{f(r)}{df(r)/dr} = \operatorname{tg}\Phi(r) \tag{A.3}$$

and Eq. (A.1) can be written in the form of the following system of first-order differencial equations:

$$\frac{d\Phi}{dr} = \cos^2\Phi + 2\left(E_{Schr} - U_{eff}^F\right)\sin^2\Phi, \tag{A.4}$$

$$\frac{d\ln P}{dr} = \left(1 - 2\left(E_{Schr} - U_{eff}^F\right)\right)\sin\Phi\cos\Phi. \tag{A.5}$$

It should be noted that Eq. (A.5) must be solved after determining eigenvalues $\varepsilon_n$ and eigenfunctions $\Phi_n(r)$ from Eq. (A.4).

For bound states with $r \to \infty$, we get

$$\operatorname{tg}\Phi\big|_{r\to\infty} = -\frac{1}{\sqrt{1-\varepsilon^2}},$$
$$\Phi\big|_{r\to\infty} = -\operatorname{arctg}\frac{1}{\sqrt{1-\varepsilon^2}} + k\pi. \tag{A.6}$$

For exponentially increasing solutions for $r \to \infty$, we have

$$\left.\text{tg}\Phi\right|_{r\to\infty} = \frac{1}{\sqrt{1-\varepsilon^2}},$$
$$\left.\Phi\right|_{r\to\infty} = \arctan\frac{1}{\sqrt{1-\varepsilon^2}} + k\pi. \quad (A.7)$$

In expressions (A.6) and (A.7) $k = 0, \pm 1, \pm 2, \ldots$

The numerical method of solving equations of type (A.4), (A.5) with asymptotic forms (A.6) and (A.7) is described in detail in [3] - [5], [26].

**Acknowledgements**

We are grateful to G.M.Ter-Akop'yan, K.O.Vlasov, M.V. Gorbatenko, V.A. Zhmailo, A.I.Milshtein, and V.E.Shemarulin for fruitful discussions. Thanks are due to A.L.Novoselova for technical support in the preparing this article.